\documentclass[11pt,a4paper]{article}
\usepackage{fullpage}
\usepackage{amsmath,amssymb,amsfonts,amsxtra,graphics,graphicx,amsthm,bbm,array}
\usepackage{hyperref}
\usepackage{CJK}

\usepackage[T1]{fontenc}
\usepackage[utf8]{inputenc}
\usepackage{authblk}
\usepackage{dcolumn}
\usepackage[vcentermath]{youngtab}
\newcommand{\mdot}{\raise1.5pt \hbox{.}}
\newcommand{\bZ}{\ensuremath{\mathbb{Z}}}
\newcommand{\bC}{\ensuremath{\mathbb{C}}}
\newcommand{\sll}{\ensuremath{\mathfrak{sl}}}
\def\bea{\begin{eqnarray}}
\def\eea{\end{eqnarray}}
\def\be{\begin{equation}}
\def\ee{\end{equation}}
\def\ba{\begin{align}}
\def\ea{\end{align}}
\begin{document}

\title{\begin{flushright}\begin{small}
{NIKHEF-2013-005}\end{small}
\end{flushright}\bigskip
Multiplicity-free quantum $6j$-symbols for $U_q(\sll_N)$}

\author[1]{Satoshi Nawata}
\author[2]{Ramadevi Pichai}
\author[2]{Zodinmawia}
\affil[1]{NIKHEF theory group, Science Park 105,1098 XG Amsterdam, The Netherlands}
\affil[2]{Department of Physics, Indian Institute of Technology Bombay, Mumbai, India, 400076}
\date{}

  \maketitle

\abstract{
We conjecture a closed form expression for the simplest class of multiplicity-free quantum $6j$-symbols for $U_q(\sll_N)$. The expression is a natural generalization of the quantum $6j$-symbols for $U_q(\sll_2)$ obtained by Kirillov and Reshetikhin. Our conjectured form enables computation of  colored HOMFLY polynomials  for various knots and links  carrying arbitrary symmetric representations.
\\}



\Yboxdim5pt
%------------------------------------

\vspace{.5cm}
\section{Introduction} From the beginning of the twentieth century, we have witnessed the mutual interactive developments between quantum physics and representation theory. Right from the birth of quantum mechanics, applications of representation theory to quantum physics have turned out to be indispensable to the study of symmetries inherent in a quantum system.  
It is well-known that the Clebsch-Gordan coefficients ($3j$-symbols) in decomposition of tensor product of irreducible representations of $\sll_2({\mathbb C})$ naturally appear in quantum theory of angular momenta.  Furthermore, in the study of atomic spectroscopy \cite{Racah:1942}, the Racah coefficients were defined as linear combinations
of products of four Clebsch-Gordan coefficients. Around the same time, by studying algebra the $3j$-symbols satisfy, Wigner independently introduced (classical) $6j$-symbols \cite{Wigner:1940,Wigner:1959}. 
The $6j$-symbols are of importance in all situations where the recoupling of angular momenta is involved.

Inspired by ideas coming from quantum physics, quantum deformation of universal enveloping algebra of a semi-simple Lie algebra, {\it a.k.a.} a quantum group, was introduced by Drinfel'd and Jimbo \cite{QG:1985,Jimbo:1985zk}, which captures the symmetry behind the Yang-Baxter equations. Meanwhile, on a separate track, Jones constructed a new polynomial invariant of links using von Neumann algebra \cite{Jones:1985dw}. It soon became apparent that the basic algebraic structures of the polynomial invariants of links could be described by quantum groups. This viewpoint leads to a wide class of polynomial invariants of links \cite{Turaev:1988eb,Reshetikhin:1988iw,Reshetikhin:1990pr}. Furthermore, quantum groups are also related to the two-dimensional Wess-Zumino-Novikov-Witten (WZNW) model. The actions of braid groups on conformal blocks of WZNW model turn out to be equivalent to the braid group representation obtained from the universal $\cal R$-matrix of the quantum group \cite{DK,Drinfeld:1989st,AlvarezGaume:1988v,AlvarezGaume:1989aq}. Hence, the polynomials invariants of links can be obtained from monodromy representations along solutions of the Knizhnik-Zamolodchikov equations.

It has been proven that the quantum deformations of the Clebsch-Gordan coefficients and the $6j$-symbols which naturally appear in the representation theory of quantum groups connect these areas of mathematics and physics in a beautiful way \cite{Kirillov:1989}. However, the generalization of the quantum $6j$-symbols for $U_q(\sll_2)$ \cite{Kirillov:1989} to higher ranks has remained a challenging open problem. In this paper, we shall conjecture a closed form expression of the simplest class of the quantum $6j$-symbols for $U_q(\sll_N)$. 
%------------------------------------
\vspace{.3cm}
\section{Quantum $6j$-symbols} Let us denote the spin-$j$ representation of $U_q(\sll_2)$ by $V_{j}$ whose highest weight is $\lambda=2j$ ($j\in\frac12\bZ$). 
The space of four-point conformal blocks is the space of linear maps ${\rm Hom}_{U_q(\sll_2)}(V_{j_1}\otimes V_{j_2}\otimes V_{j_3}\otimes V_{j_4}^*,\bC)$ invariant under the diagonal action of $U_q(\sll_2)$ and, most importantly, it is a \emph{finite-dimensional} vector space. By associativity of the tensor product, we can take a basis in two ways;
\bea
{\rm Hom}_{U_q(\sll_2)}\left( (V_{j_1}\otimes V_{j_2})_{j_{12}}\otimes V_{j_3},V_{j_4}\right)=\bigoplus_{j_{12}} \bC {\raisebox{-.6cm}{\includegraphics[width=1.8cm]{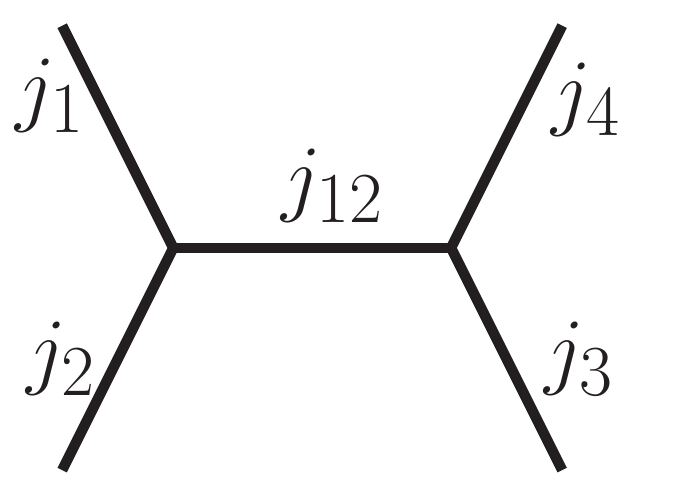}}}, \ \ \ \
\eea
and 
\bea
{\rm Hom}_{U_q(\sll_2)}\left( V_{j_1}\otimes (V_{j_2}\otimes V_{j_3})_{j_{23}}, V_{j_4}\right)=\bigoplus_{j_{23}} \bC {\raisebox{-.7cm}{\includegraphics[width=1.8cm]{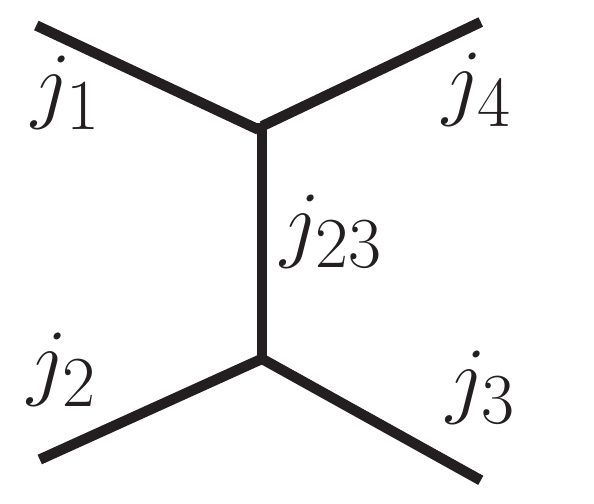}}},\ \ \ \
\eea
where $j_{12}$ and $j_{23}$ satisfy the quantum Clebsch-Gordan condition (the fusion rule); $j_1+j_2+j_{12}\in \bZ$ and $\vert j_1-j_2\vert \le j_{12} \le j_1+j_2$. (The same condition for $j_{23}$.) Then, the quantum $6j$-symbols $\left\{ \begin{matrix}
j_1 & j_2 & j_{12} \\ j_3 & j_4 & j_{23} \end{matrix} \right\}$ for $U_q(\sll_2)$ appear in the transformation matrix of the two bases
\begin{equation}
{\raisebox{-.6cm}{\includegraphics[width=2cm]{CB1}}}=     \sum_{j_{23}} a_{j_{12}\, j_{23}} 
\left[ \begin{matrix} j_1 & j_2 \\ j_3 & j_4\end{matrix} \right]{\raisebox{-.8cm}{\includegraphics[width=2cm]{CB2}}}~,
\label{racah}
\end{equation}
where \bea
&&a_{j_{12}\, j_{23}}\left[ \begin{matrix} j_1 & j_2 \\ j_3 & j_4\end{matrix} \right]=(-1)^{j_1+j_2+j_3+j_4} \sqrt{ [2j_{12}+1][2j_{23}+1]} 
\left\{ \begin{matrix}
j_1 & j_2 & j_{12} \\ j_3 & j_4 & j_{23} \end{matrix} \right\}~. \ \ \ \ \ 
\label{wig}
\eea
Here, the square bracket  defines a $q$-number  
$$[n] =\frac{q^{n / 2}-q^{-n / 2}}{ q^{1 / 2}-q^{-1 / 2}}~,$$ 
and the transformation matrix $a_{j_{12}\, j_{23}}$ is usually called the fusion matrix. The rigorous derivation of a closed form expression of the quantum $6j$-symbols for $U_q(\sll_2)$ was first given by Kirillov and Reshetikhin \cite{Kirillov:1989}. Later,  Masbaum and Vogel provided another derivation based on linear skein theory \cite{Masbaum}.

The generalization to higher ranks requires replacement of a spin $j$ by a highest weight $\lambda$ of a representation of $U_q(\sll_N)$. A representation of $U_q(\sll_N)$ with a highest weight $\lambda$ can be equivalently specified by a Young tableau $\{\ell_i\}_{1\le i \le N-1}$ with $\ell_1\ge \cdots\ge \ell_{N-1}$. If one writes the highest weight as $\lambda=\sum_{i=1}^{N-1}\lambda_i\omega_i$ where $\{\omega_i\}_{1\le i \le N-1}$ are the fundamental weights, the relation to the Young tableau can be read off by $\ell_i=\lambda_i +\lambda_{i+1}+\cdots +\lambda_{N-1}$. In what follows, we identify a highest weight with a Young tableau by this dictionary. For general $N$, it is necessary to introduce the conjugate representation $V_{\lambda^*}$ of the representation with the highest weight $\lambda$ where $\lambda^*:=\sum_{i=1}^{N-1} \lambda_{N-i}\omega_i$. Then, the fusion rule of quantum $6j$-symbols for $U_q(\sll_N)$ 
\be
\left\{ \begin{matrix} \lambda_1 & \lambda_2 & \lambda_{12} \\ \lambda_3 & \lambda_4 & \lambda_{23}\end{matrix} \right\} 
\ee
is that $V_{\lambda_{12}}\in (V_{\lambda_{1}} \otimes V_{\lambda_{2}}) \cap (V_{\lambda_{3}^*}\otimes V_{\lambda_{4}^*})$ and $V_{\lambda_{23}}\in (V_{\lambda_{2}} \otimes V_{\lambda_{3}} )\cap (V_{\lambda_{1}^*} \otimes V_{\lambda_{4}^*})$.
From the construction, we expect the quantum $6j$-symbols to satisfy the following symmetries:
\bea\label{symmetry}
\left\{ \begin{matrix} \lambda_1 & \lambda_2 & \lambda_{12} \\ \lambda_3 & \lambda_4 & \lambda_{23}\end{matrix} \right\} &=& \left\{ \begin{matrix} \lambda_3 & \lambda_2 & \lambda_{23}
\\ \lambda_1 & \lambda_4 & \lambda_{12}\end{matrix} \right\}  = \left\{ \begin{matrix} \lambda_1^* & \lambda_2^* & \lambda_{12}^* \\ \lambda_3^* & \lambda_4^* & \lambda_{23}^*\end{matrix} \right\}  \cr
&=& \left\{ \begin{matrix}
\lambda_1 & \lambda_4 & \lambda^*_{23} \cr \lambda_3 & \lambda_2 & \lambda^*_{12}\end{matrix} \right\}  =\left\{ \begin{matrix} \lambda_2 & \lambda_1 & \lambda_{12} \\ \lambda_4 & \lambda_3 & \lambda_{23}^*\end{matrix} \right\}
\label{A.4}
\eea
In addition, the relationship between the fusion matrix and the quantum  $6j$-symbols for $U_q(\sll_N)$ is generalized to 
\bea
&&a_{\lambda_{12}\lambda_{23}}\left[ \begin{matrix} \lambda_1 & \lambda_2 \\  \lambda_3& {\lambda_4}\end{matrix} \right] =
\epsilon_{\{\lambda_i\}} \sqrt{\dim_q V_{\lambda_{12}}\dim_q V_{\lambda_{23}}} \left\{ \begin{matrix}
\lambda_{1} & \lambda_{2 }& \lambda_{{12}} \\ \lambda_{3} & \lambda_{4} & \lambda_{{23}} \end{matrix} \right\}~,
\label{A.1}
\eea
where $\epsilon_{\{\lambda_i\}}=\pm 1$ and $\dim_q V_\lambda$ is the quantum dimension of the representation $V_\lambda$ with highest weight $\lambda$.

Unlike the representations of $U_q(\sll_2)$, there are serious technical difficulties for $U_q(\sll_{N})$ in the fact that  the decompositions of tensor products involve multiplicity structure in general. Specifically, isomorphic irreducible constituents will arise more than once in the decomposition of a tensor product. However, there are special cases which decompose in a multiplicity-free way. Among them, we shall estrict ourselves to the simplest class which can be regarded as the natural extension of the case for $U_q(\sll_2)$ \cite{Kirillov:1989}: the tensor products of two symmetric representations, and the tensor products of a symmetric representation and a representation conjugate to a symmetric representation.  To obey the fusion rule,  two of $\lambda_1,\lambda_2,\lambda_3,\lambda_4$ need to be symmetric representations, and the other two must be conjugate to symmetric representations.
Using the symmetries \eqref{sym}, it turns out that multiplicity-free quantum $6j$-symbols for  $U_q(\sll_{N})$ of this class amount to the following two types:
\begin{itemize}
\item Type I
\bea
&&\left\{\begin{array}{ccc} n_1\omega_1& n_2 \omega_{N-1} & (n_1-n_2+k_1)\omega_1+k_1 \omega_{N-1}\\ n_3\omega_1&n_4 \omega_{N-1} & (n_3-n_2+k_2)\omega_1+k_2 \omega_{N-1}\end{array}\right\} \cr
&=&\left\{{\raisebox{-.6cm}{\includegraphics[width=6.6cm]{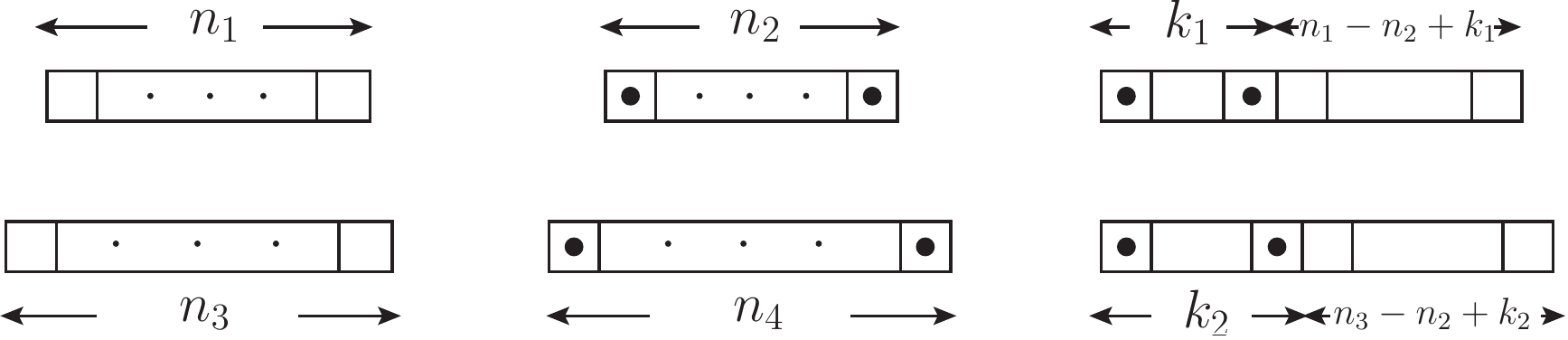}}}\right\} \ \ \
\eea
where $n_2\le n_1\le n_3$, $k_1\le n_2$ and $k_2\le n_1$. The fusion rule requires $n_1+n_3=n_2+n_4$. 

\item Type II
\bea
&&\left\{\begin{array}{ccc} n_1 \omega_1& n_2 \omega_1& (n_1+n_2-2k_1)\omega_1+k_1\omega_2\\ n_3\omega_{N-1}& n_4\omega_{N-1}& (n_2-n_3+k_2)\omega_1+k_2\omega_{N-1}\end{array}\right\}\cr
&=&\left\{{\raisebox{-.6cm}{\includegraphics[width=6.6cm]{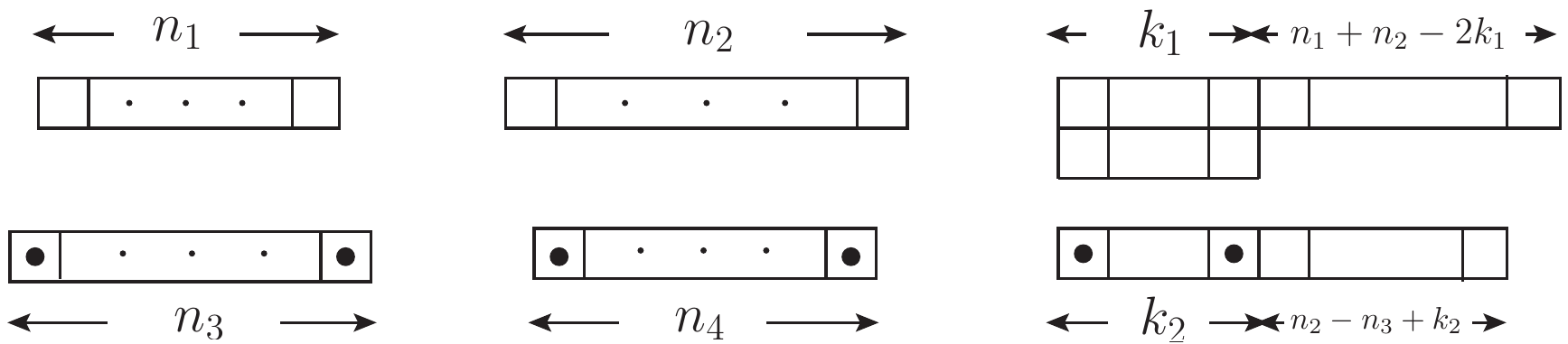}}}\right\}
\eea
where $n_1 \le n_2$, $k_2\le \min(n_1,n_3)$ and $k_1\le \min(n_1,n_3,n_4)$. The fusion rule requires $n_1+n_2=n_3+n_4$.
\end{itemize}
We note that the highest weight of the symmetric representation $\raisebox{-.1cm}{\includegraphics[width=1.4cm]{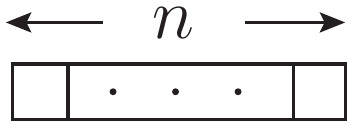}}$ is $n\omega_1$ and that of the conjugate representation $\raisebox{-.1cm}{\includegraphics[width=1.4cm]{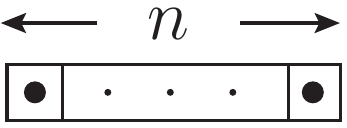}}$  is $n\omega_{N-1}$ where $\raisebox{-.05cm}{\includegraphics[width=.3cm]{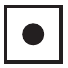}}$ represents $(N-1)$ vertical boxes $\raisebox{-.55cm}{\includegraphics[width=1.7cm]{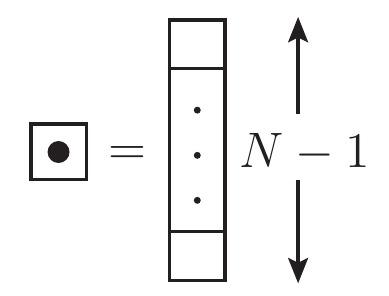}}$. 

For low representations ($n_i \leq 2$) we have determined the quantum $6j$-symbols
\cite{Ramadevi,Zodinmawia:2011oya}.  Extension to higher representations ($n_i \leq 4$)  is achieved 
through the following route:  In our recent paper \cite{Nawata:2012pg}, we have  
colored HOMFLY polynomials for a class of knots called twist knots $K_p$ where $p$ 
denotes the number of full-twists.  Using the identities obeyed by quantum 6j symbols and  
equating the colored HOMFLY polynomials of these
twist knots  with the $SU(N)$ Chern-Simons invariant involving quantum 6j-symbols 
(see (C.1)  in \cite{Nawata:2012pg}), we determined  the quantum 6j symbols for $n_i =3$ and $n_i=4$. We recognized a pattern
from the data on quantum 6j symbols ($n_i \leq 4$) motivating us to 
attempt a closed form expression for arbitrary $n_i$'s. 
 
{\bf Conjecture}
The  closed form expression for quantum 6j-symbols of type I and type II
is as follows:
\bea\label{closed}
\left\{ \begin{matrix} \lambda_{1} & \lambda_{2} & \lambda_{{12}} \\ \lambda_{3} & 
\lambda_{4} & \lambda_{{23}}\end{matrix} \right\} &=& \Delta
(\lambda_{1},\lambda_{2}, \lambda_{{12}}) \Delta (\lambda_{3},\lambda_{4},\lambda_{{12}})   \Delta (\lambda_{1},\lambda_{4},\lambda_{{23}}) \Delta
(\lambda_{2},\lambda_{3},\lambda_{{23}})\cr
&&\times  [N-1]!   \sum_{z \in \bZ_{\ge0}} (-)^z [z+N-1]!C_z(\{\lambda_{i}\},\lambda_{{12}},\lambda_{{23}})\cr
&&\times \Big\{  \left[z-\tfrac12\langle\lambda_{1} + \lambda_{2} +
\lambda_{{12}}, \alpha_1^{\vee}+\alpha_{N-1}^{\vee}\rangle\right]! \cr
&&\ \ \times \left[z -\tfrac12\langle \lambda_{3} + \lambda_{4} + \lambda_{{12}}, \alpha_1^{\vee}+\alpha_{N-1}^{\vee}\rangle\right]! \cr
&&\ \ \times  \left[z - \tfrac12\langle\lambda_{1} + \lambda_{4 }+ \lambda_{{23}}, \alpha_1^{\vee}+\alpha_{N-1}^{\vee}\rangle\right]!\cr
&&\ \ \times \left[z -\tfrac12\langle \lambda_{2} + \lambda_{3} + \lambda_{{23}}, \alpha_1^{\vee}+\alpha_{N-1}^{\vee}\rangle\right]!\cr
&&\ \ \times  \left[\tfrac12\langle\lambda_{1} + \lambda_{2 }+ \lambda_{3} + \lambda_{4},\alpha_1^{\vee}+\alpha_{N-1}^{\vee}\rangle - z\right]!
\cr
&&\ \ \times  \left[\tfrac12\langle\lambda_{1}+\lambda_{3}+\lambda_{{12}}+
\lambda_{{23}},\alpha_1^{\vee}+\alpha_{N-1}^{\vee}\rangle-z\right]! \cr
&&\ \ \times \left[\tfrac12\langle\lambda_{2}+\lambda_{4}+\lambda_{{12}}+
\lambda_{{23}}, \alpha_1^{\vee}+\alpha_{N-1}^{\vee}\rangle-z\right]!
\Big\}^{-1},\ \ \ \ \ \
\eea
where
\bea\label{delta}
\Delta(\lambda_{1},\lambda_{2},\lambda_{3}) 
&=& \left\{[\left[\tfrac12\langle-\lambda_{1}+\lambda_{2}+
\lambda_{3}, \alpha_1^{\vee}+\alpha_{N-1}^{\vee}\rangle\right]! \right.\cr
&&\times\left[\tfrac12\langle\lambda_{1}-\lambda_{2}+\lambda_{3}, \alpha_1^{\vee}+\alpha_{N-1}^{\vee}\rangle\right]! \cr
&&\times\left.\left[\tfrac12\langle\lambda_{1}+\lambda_{2}-\lambda_{3}, \alpha_1^{\vee}+\alpha_{N-1}^{\vee}\rangle\right]! \right\}^{1/2} \cr
&&\times\left\{\left[\tfrac12\langle\lambda_{1}+\lambda_{2}+\lambda_{3},\alpha_1^{\vee}+\alpha_{N-1}^{\vee}\rangle+N-1\right]!\right\}^{-1/2}~. \ \ \ \ \ \
\eea
Here, we define the $q$-factorial by $[n]!=[n][n-1]\cdots[3][2][1]$
and take $[0]!=1$. We use the fact that $\alpha_j^{\vee}$ are duals of simple roots which form a basis of coroots, and the paring with the fundamental weights provides $\langle \omega_i, \alpha_j^{\vee}\rangle = \delta_{ij}$. 

It is important to stress that only finite number of positive integers $z$ in the summation 
will give non-zero  contribution. The factors $C_z(\{\lambda_{i}\}, \lambda_{{12}},\lambda_{{23}})$ in \eqref{closed} for type I are
\bea
&C_z\left( \left\{\begin{array}{ccc} n_1\omega_1& n_2 \omega_{N-1} & (n_1-n_2+k_1)\omega_1+k_1 \omega_{N-1}\\ n_3\omega_1&n_4 \omega_{N-1} & (n_3-n_2+k_2)\omega_1+k_2 \omega_{N-1}\end{array}\right\} \right)\cr
&=\left\{\begin{array}{ll}\delta_{z,z_{\min}+i} \left[\begin{array}{c}  N-2+k_2-i\\ k_2-i \end{array} \right]^{-1} &\quad {\rm for} \ k_1>k_2 \,,\\ \delta_{z,z_{\min}+i} \left[\begin{array}{c}  N-2+k_1-i\\ k_1-i \end{array} \right]^{-1} &\quad {\rm for} \ k_1\le k_2 \,,\end{array}\right.
\eea
where $z_{\rm min}$ is the smallest integer $z$ in the summation in \eqref{closed} which gives a non-trivial value.
Similarly, for type II, the factors $C_z(\{\lambda_{i}\}, \lambda_{{12}},\lambda_{{23}})$ are
\bea
&C_z\begin{small}\left(\left\{\begin{array}{ccc} n_1 \omega_1& n_2 \omega_1& (n_1+n_2-2k_1)\omega_1+k_1\omega_2\\ n_3\omega_{N-1}& n_4\omega_{N-1}& (n_2-n_3+k_2)\omega_1+k_2\omega_{N-1}\end{array}\right\}\right)\end{small}\cr
&=\left\{\begin{array}{ll}\delta_{z,z_{\max}-i} \left[\begin{array}{c}  N-2+k_2-i\\ k_2-i \end{array} \right]^{-1} &\quad {\rm for} \ k_1>k_2\,, \\ \delta_{z,z_{\max}-i} \left[\begin{array}{c}  N-2+k_1-i\\ k_1-i \end{array} \right]^{-1} &\quad {\rm for} \ k_1\le k_2\,, \end{array}\right.
\eea
where $z_{\rm max}$ is the largest integer $z$ in the summation in \eqref{closed} which gives a non-trivial value.
We denote the $q$-binomial by
\bea
\left[ \begin{array}{c} p \\ q   \end{array} \right] =\frac{[p]!}{[q]![p-q]!}~.
\eea
Although there are the square roots in the expression \eqref{delta}, the $6j$-symbols \eqref{closed} are actually rational functions with respect to $q^{1/2}$. Obviously, it is easy to see that the expression \eqref{closed} reduces to the form of $U_q(\sll_2)$ provided by Kirillov and Reshetikhin \cite{Kirillov:1989} when we take $N=2$.  

In addition, we have checked that \eqref{closed} satisfy the orthogonal property
\bea
&&\sum_{\lambda_{12}} \dim_qV_{\lambda_{12}}\dim_qV_{\lambda_{23}} \left\{ \begin{matrix} \lambda_1 & \lambda_2 & \lambda_{12} \\ \lambda_3 & \lambda_4 & \lambda_{23}
\end{matrix}\right\}  \left\{ \begin{matrix} \lambda_1 & \lambda_2 & \lambda_{12} \cr \lambda_3 & \lambda_4 & \lambda'_{23} \end{matrix} \right\}\cr
&&  \hspace{4cm}=
\delta_{\lambda_{23}\lambda'_{23}}~, 
\label{A.6}
\eea
and the Racah identity
\bea
&&\sum_{\lambda_{12}} \epsilon_{\{\lambda_{12},\lambda_{23},\lambda_{24}\}} q^{-\frac{C_{12}}{2}} \dim_qV_{\lambda_{12}} \left\{ \begin{matrix} \lambda_1 & \lambda_2 & \lambda_{12} \\ \lambda_3
& \lambda_4 & \lambda_{23}\end{matrix} \right\}  \left\{ \begin{matrix} \lambda_1 & \lambda_2 & \lambda_{12} \\ \lambda_4 & \lambda_3 & \lambda_{24}\end{matrix} \right\}
\cr
&&  =   \left\{ \begin{matrix} \lambda_3 & \lambda_2 & \lambda_{23} \\ \lambda_4 & \lambda_1 & \lambda_{24}\end{matrix} \right\}
q^{\frac{C_{23} + C_{24}}{2}}
q^{-\frac{C_{1} + C_{2} +C_{3} + C_{4}}{2}} ~,
\label{A.7}
\eea
as well as the identity
\be
\left\{ \begin{matrix} \lambda_1 & \lambda_2 & 0 \\ \lambda_3 & \lambda_4 & \lambda_{23}\end{matrix} \right\} =
\frac{\epsilon_{\{\lambda_2,\lambda_3,\lambda_{23}\}}
\delta_{\lambda_1\lambda_2} \delta_{\lambda_3\lambda_4}}{\sqrt{\dim_qV_{\lambda_2} \dim_qV_{\lambda_3}}} ~.
\label{A.5}
\ee
Here $C_i$ is the quadratic Casimir invariant for the representation of  highest weight $\lambda_i$.
The sign $\epsilon=\pm1$ can be easily read off by comparing it with the results for $U_q(\sll_2)$.

 We expect that the quantum $6j$-symbols for $U_q(\sll_N)$ obey the pentagon (Biedenharn-Elliot) identity
\bea
&&\sum_{\mu_1}\epsilon_{\{\mu_i,\kappa_i\}} \dim_qV_{\mu_1}\left\{
\begin{matrix}\kappa_1 & \lambda_3 & \kappa_2 \\ \lambda_4 & \lambda_5 & \mu_1 \end{matrix} \right\}
\quad \left\{ \begin{matrix} \lambda_1 & \lambda_2 & \kappa_1 \\ \mu_1 & \lambda_5 & \mu_2 \end{matrix} \right\} \quad
\left\{ \begin{matrix} \mu_2 & \lambda_2 & \mu_1 \cr \lambda_3 & \lambda_4 & \mu_3 \end{matrix} \right\}\cr
&& \ = \ \epsilon_{\{\lambda_i\}}
\left\{ \begin{matrix} \lambda_1 & \mu_3 & \kappa_2 \cr \lambda_4 & \lambda_5 & \mu_2 \end{matrix} \right\} \quad
\left\{ \begin{matrix}\lambda_1 & \lambda_2 & \kappa_1 \\ \lambda_3 & \kappa_2 & \mu_3 \end{matrix} \right\}  ~. 
\label{A.8}
\eea
However, this property cannot be checked unless expressions beyond the multiplicity-free ones are obtained.

With our conjectured quantum $6j$-symbols \eqref{closed}, we can verify that they 
reproduce the  HOMFLY polynomials colored by the symmetric representation of many non-torus knots  
 \cite{Itoyama,Itoyama:2012re}, the Whitehead link and the Borromean rings \cite{Kawagoe,Gukov(to appear)} up to 4 boxes.  Moreover, we compute colored HOMFLY polynomials of many knots and links which we tabulate in the companion paper \cite{HOMFLY:2013}. However, 
it is not clear, at present, whether the proof \cite{Kirillov:1989,Masbaum} can be extended to our conjectured  $sl_N$ quantum
6j symbols mainly due to the presence of $C_z(\{\lambda_{i}\}, \lambda_{{12}},\lambda_{{23}})$ in \eqref{closed}.

\vspace{.3cm}

\section{Discussion}
In this paper, we proposed the closed form expression of the quantum $6j$-symbols for $U_q(\sll_N)$. However, the structure behind $6j$-symbols is by far richer and we only scratch the surface of this topic. Firstly,  a rigorous derivation of \eqref{closed} by quantum groups still remain an open problem. In addition to this, further study needs to be undertaken to obtain the expressions for other multiplicity-free cases as well as general cases. For this purpose, it is necessary to investigate explicit expressions for the quantum $3j$-symbols and their relation to the quantum $6j$-symbols.

There is an important property of quantum $6j$-symbols for $U_q(\sll_N)$ which will be useful for evaluating quantum $6j$-symbols beyond symmetric representations. We can observe that a quantum $6j$-symbol involving anti-symmetric representations and their conjugate representations can be obtained by changing $[N+k] \to [N-k]$ in the quantum $6j$-symbol for symmetric representations related by transposition (mirror reflection across the diagonal). For example, we can find the following $6j$-symbols using properties \eqref{A.6}, \eqref{A.7} and \eqref{A.5}:
\bea\label{antisym}
\left\{{\raisebox{-1.05cm}{\includegraphics[width=2.6cm]{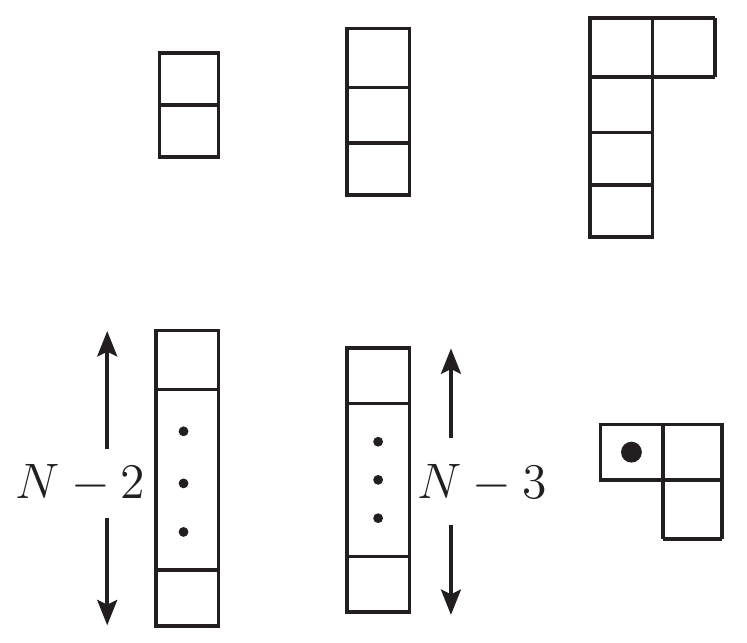}}}\right\}=\tfrac{[2]^2}{[N+1][N][N-1][N-2]^2}\left(\tfrac{[3][N-3][N-4]-[N][N+1]}{[N]+[3][N-4]}\right)~. \ \
\eea
The quantum $6j$-symbol related by transposition is 
\bea\label{sym}
\left\{{\raisebox{-.4cm}{\includegraphics[width=3.2cm]{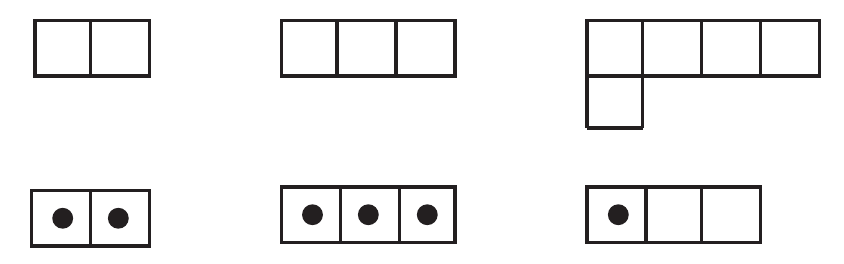}}}\right\}=\tfrac{[2]^2}{[N-1][N][N+1][N+2]^2}\left(\tfrac{[3][N+3][N+4]-[N][N-1]}{[N]+[3][N+4]}\right)~.\ \
\eea
It is easy to see the symmetry $[N+k] \leftrightarrow [N-k]$  
between the two quantum $6j$-symbols.
In fact, this explains the symmetry between the colored HOMFLY polynomials colored by symmetric representation $S^r$ and anti-symmetric representations $\Lambda^r$:
\bea\label{mirror}
P_{S^r}(K;a,q)=P_{\Lambda^r}(K;a,q^{-1})~.
\eea
Following \cite{Gukov:2011ry}, we call this property the \emph{mirror symmetry} of quantum $6j$-symbols.
In this way, every multiplicity-free quantum $6j$-symbol involving anti-symmetric representations can be obtained from its mirror dual $6j$-symbol which can be explicitly evaluated by \eqref{closed}. Nevertheless, a closed form expression using highest weights as in \eqref{closed} is still lacking in anti-symmetric representations. Certainly, it would be intriguing to see whether the mirror symmetry holds beyond multiplicity-free quantum $6j$-symbols.

Another important aspect of quantum $6j$-symbols is their relationship with  $q$-hypergeometric functions \cite{Askey:1979}. The quantum $6j$-symbols for $U_q(\sll_2)$ can be expressed as the balanced hypergeometric function ${}_4\phi_3$ \cite{Kirillov:1989}. However,  for $U_q(\sll_N)$, the coefficients $C_z(\{\lambda_i\},\lambda_{12},\lambda_{23})$ prevents us from writing the expression \eqref{closed} in terms of the balanced hypergeometric function ${}_4\phi_3$. Therefore, it is important to study the connection to generalized $q$-hypergeometric functions \cite{Askey:1979}. Besides, it is well-known that there are many different ways to express the quantum $6j$-symbols for $U_q(\sll_2)$. Hence, it would be worthwhile to find the other expressions for the quantum $6j$-symbols for $U_q(\sll_N)$.

Furthermore, there is a geometric interpretation of the quantum $6j$-symbols.
One can associate the quantum $6j$-symbols to a tetrahedron whose edges are colored by representations of $U_q(\sll_N)$  \cite{Turaev:1992hq}. Although it is necessary to have quantum $6j$-symbols for arbitrary representations to obtain invariants  of 3-manifolds \cite{Turaev:1992hq} in the context of $U_q(\sll_N)$, the expression \eqref{closed} is suitable to study the large color behavior \cite{Ponzano}. 
Therefore, it would be interesting to explore the large color behavior of quantum $6j$-symbols and their relation to the geometry of  the complement of a tetrahedron in $S^3$.

As we have seen, quantum $6j$-symbols are very interesting in their own right and contain remarkable mathematical structure. Despite  their long history, they are indeed among the least understood quantities in mathematical physics.
While in this paper we focus on the simplest class of multiplicity-free quantum $6j$-symbols for $U_q(\sll_N)$, we hope
that our results will serve as a stepping stone towards the study of general quantum $6j$-symbols.

%------------------------------------
\vspace{.3cm}
{\bf Acknowledgements}:
The authors are indebted to Stavros Garoufalidis and Jasper Stokman for the useful discussions and comments. S.N. and Z. would like to thank Indian String Meeting 2012 at Puri for providing a stimulating academic environment. S.N. is grateful to IIT Bombay  for its warm hospitality.  The work of S.N. is partially supported by the ERC Advanced
Grant no.~246974, {\it "Supersymmetry: a window to non-perturbative physics"}.

\bibliographystyle{alpha}

\begin{thebibliography}{9}

\bibitem{AlvarezGaume:1988v}
Alvarez-Gaume, L., Gomez, C., Sierra, G.: Quantum Group Interpretation of Some Conformal Field Theories. \href{http://link.aps.org/doi/10.1016/0370-2693(89)90027-0}{Phys.Lett. {\bf B220} 142 (1989)}

\bibitem{AlvarezGaume:1989aq}
Alvarez-Gaume, L., Gomez, C., Sierra, G.: Duality and Quantum Groups. \href{http://link.aps.org/doi/10.1016/0550-3213(90)90116-U}{Nucl.Phys. {\bf B330} 347 (1990)}

\bibitem{Askey:1979}
Askey, R., Wilson, J.: A set of orthogonal polynomials that generalize the Racah coefficients or 6-j symbols. \href{http://dx.doi.org/10.1137/0510092}{SIAM J. Math. Anal. {\bf 10} no.5 1008 (1979)}

\bibitem{QG:1985}
 Drinfel'd, V.G.: Hopf algebras and the quantum Yang-Baxter equation. Sov.Math.Dokl. {\bf 32} 254-258 (1985)
 
\bibitem{Drinfeld:1989st}
Drinfel'd, V.G.: Quasi-hopf algebras. Leningrad Math. J. {\bf 1} 1419-1457 (1990)


\bibitem{Gukov:2011ry}
Gukov, S., Stosic, M.: Homological Algebra of Knots and BPS States. \href{http://arxiv.org/abs/1112.0030}{arXiv:1112.0030}

\bibitem{Gukov(to appear)}
Gukov, S., Nawata, S., Stosic, M., Su{\l}kowski, P.: (2013, in preparation)


\bibitem{Itoyama}
Itoyama, H., Mironov, A., Morozov, A., Morozov, An.: Character expansion for HOMFLY polynomials. III. All 3-Strand braids in the first symmetric representation. \href{http://dx.doi.org/10.1142/S0217751X12500996}{Int. J. Mod. Phys. {\bf A27} (2012) 1250099}

\bibitem{Itoyama:2012re}
Itoyama, H., Mironov, A., Morozov, A., Morozov, An.: Eigenvalue hypothesis for Racah matrices and HOMFLY polynomials for 3-strand knots in any symmetric and  antisymmetric representations. \href{http://dx.doi.org/10.1142/S0217751X13400095 }{Int. J. Mod. Phys. {\bf A28} (2013) 1340009}

\bibitem{Jimbo:1985zk}
Jimbo, M.: A $q$-difference analogue of $U(g)$ and the Yang-Baxter equation. \href{http://dx.doi.org/10.1007/BF00704588}{Lett. Math. Phys. {\bf 10} 63-69 (1985)}

\bibitem{Jones:1985dw}
Jones, V.F.R.: A polynomial invariant for knots via von Neumann algebras. \href{http://dx.doi.org/10.1090/S0273-0979-1985-15304-2 }{Bull. Am. Math. Soc. {\bf 12} 103-111 (1985)}

\bibitem{Kawagoe}
Kawagoe, K.: 
On the formulae for the colored HOMFLY polynomials. 
\href{http://arxiv.org/abs/1210.7574}{arXiv:1210.7574}


\bibitem{Kirillov:1989}
Kirillov, A.N., Reshetikhin, N.Yu.: Representations of the algebra $U_q(sl(2))$, q-orthogonal polynomials and invariants of Links. In: Kohno, T. (ed.) New Developments in the Theory of Knots. World Scientific, Singapore (1989)

\bibitem{DK}
Kohno, T.: Monodromy representations of braid groups and Yang-Baxter equations. \href{http://link.aps.org/doi/10.5802/aif.1114 }{Ann. Inst. Fourier {\bf 37} no.4 139-160 (1987)}

\bibitem{Masbaum}
 Masbaum, G., Vogel, P.: 3-valent graphs and the Kauffman bracket. \href{http://projecteuclid.org/euclid.pjm/1102622100}{Pac. J. Math 164.2 (1994): 361-381}
 
 \bibitem{HOMFLY:2013}
Nawata, S., Ramadevi, P., Zodinmawia: Colored HOMFLY polynomials from Chern-Simons theory. \href{http://arxiv.org/abs/1302.5144}{arXiv:1302.5144}

\bibitem{Nawata:2012pg}
Nawata, S., Ramadevi, P., Zodinmawia, Sun, X.: Super-A-polynomials for Twist Knots.
\href{http://dx.doi.org/10.1007/JHEP11(2012)157}{JHEP {\bf 1211}, 157 (2012)}.
\href{http://arxiv.org/abs/1209.1409}{arXiv:1209.1409}

\bibitem{Ponzano}
Ponzano, G., Regge, T.: Semiclassical limit of Racah coefficients. In: Spectroscopy and Group Theoretical Methods in Physics, Amsterdam, pp. 1--58

\bibitem{Racah:1942}
Racah, G.: Theory of Complex Spectra. II. \href{http://link.aps.org/doi/10.1103/PhysRev.62.438}{Phys. Rev. {\bf 62} 438-462 (1942)}

\bibitem{Ramadevi}
Ramadevi, P., Govindarajan, T.R., Kaul, R.K.:  Three-dimensional Chern-Simons theory as a theory of knots and links. III Compact semisimple group. \href{http://dx.doi.org/10.1016/0550-3213(93)90652-6}{Nucl. Phys. {\bf B402}, 548--566 (1993)}. \href{http://arxiv.org/abs/hep-th/9212110}{hep-th/9212110}

\bibitem{Reshetikhin:1988iw}
Reshetikhin, N.Yu.: Quantized universal enveloping algebras, the Yang-Baxter equation and invariants of links I, II, LOMI preprints E-4-87 \& E-17-87, Leningrad (1988)

\bibitem{Reshetikhin:1990pr}
Reshetikhin, N.Yu., Turaev, V.G.: Ribbon graphs and their invariants derived from quantum groups. \href{http://dx.doi.org/10.1007/BF02096491}{Commun. Math. Phys. {\bf 127} 1-26 (1990)}
 
\bibitem{Turaev:1988eb}
Turaev, V.G.: The Yang-Baxter equation and invariants of links. \href{http://dx.doi.org/10.1007/BF01393746}{Invent. Math. 92 527-553 (1988)}

\bibitem{Turaev:1992hq}
Turaev, V., Viro, O.: State sum invariants of 3-manifolds and quantum 6j-symbols. \href{http://dx.doi.org/10.1016/0040-9383(92)90015-A}{Topology 31 865-902 (1992)}

\bibitem{Wigner:1940}
Wigner, E.P.: Manuscript in 1940, appeared in Quantum Theory of Angular Momentum, pp. 87--133. (Academic Press, New York 1965)

\bibitem{Wigner:1959}
Wigner, E.P.: Group Theory and Its Application to the Quantum Mechanics of Atomic Spectra (Academic Press, New York 1959)

\bibitem{Zodinmawia:2011oya}
Zodinmawia, Ramadevi, P.: SU(N) quantum Racah coefficients and non-torus links. \href{http://dx.doi.org/10.1016/j.nuclphysb.2012.12.020}{Nucl. Phys. B {\bf 870}, 205--242 (2013)}. \href{http://arxiv.org/abs/1107.3918}{arXiv:1107.3918}

%
%\bibitem{Physics}
%G.W.~Moore and N. Seiberg, \href{http://dx.doi.org/10.1016/0370-2693(88)91796-0}{Phys.Lett. {\bf B212} (1988) 451
%G.W.~Moore and N. Seiberg, \href{http://dx.doi.org/10.1007/BF01238857}{Commun.Math.Phys. {\bf 23} (1989) 177}
%E.~Witten, \href{http://dx.doi.org/10.1007/BF01217730}{Commun. Math. Phys. {\bf} (1989) 351-399} 
\end{thebibliography}

\end{document}